\documentclass[12pt]{article}
\usepackage{graphics}

\def\putgraph#1#2{
    \vskip 0.5cm 
    \centerline{\resizebox{#1}{!}{\includegraphics{#2}}}
}

\def\p{\partial}
\def\cc{\hbox{c.c.}}
\def\Krp{{\bf Q'}\cdot{\bf r}'}
\def\kr{{\bf q}\cdot{\bf r}}
\def\Kr{{\bf Q}\cdot{\bf r}}

\begin{document}

\title{Lamellar phase stability in diblock copolymers under reciprocating 
shear flows}
\author{Peilong Chen \\
Department of Physics \\
National Central University \\
Chungli 320, Taiwan \\
\\
and \\
\\
Jorge Vi\~nals$^*$ \\
School of Computational Science and Information Technology \\
Florida State University, Tallahassee, Florida 32306-4120 \\
and Department of Chemical Engineering, \\
FAMU-FSU College of Engineering, Tallahassee, Florida 32310-6046}

\maketitle

\begin{abstract}

A mesoscopic model of a diblock copolymer is used to study the
stability of a uniform lamellar phase under a reciprocating shear
flow. Approximate viscosity contrast between the microphases is allowed 
through a linear dependence of the (Newtonian) shear viscosity on monomer
composition. We first show that viscosity contrast does not affect the 
composition of the base lamellar phase in an unbounded geometry, and that it
only couples weakly to long wavelength perturbations. A perturbative analysis
is then presented to address the stability of uniform lamellar structures
under long wavelength perturbations by self-consistently solving for 
the composition and velocity fields of the perturbations. Stability 
boundaries are obtained as functions of the physical parameters of 
the polymer, the parameters of the flow and the initial orientation
of the lamellae. We find that all orientations are linearly stable
within specific ranges of parameters, but that the perpendicular
orientation is generally stable within a larger range than the parallel 
orientation.
Secondary instabilities are both of the Eckhaus type (longitudinal) and 
zig-zag type (transverse). The former is not expected to lead to 
re-orientation of the lamella, whereas in the second case the critical 
wavenumber is typically found to be along the perpendicular orientation.

\end{abstract}

\section{Introduction}
\label{sec:introduction}

Diblock copolymers are macromolecules comprising two chemically
distinct and mutually incompatible segments (monomers) that are
covalently bonded.  The equilibrium properties are determined by the
degree of polymerization $N$, (i.e., the length of the polymer chain),
the volume fraction of one of the monomers $f$, and the Flory-Huggins
interaction parameter between the distinct segments $\chi$
\cite{re:bates90,re:hamley98}. While the degree of polymerization and
the monomer volume fraction are determined by the processing
conditions, the value of the parameter $\chi$ is entirely determined
by the choice of monomers and temperature.

Above the order-disorder transition temperature $T_{ODT}$, the
equilibrium phase is disordered and the monomer concentration uniform.
In mean field theory, the order-disorder transition takes place at
$\chi N \simeq 10$, where the Flory-Huggins parameter and temperature
$T$ are related through $\chi = \alpha/T + \beta$, where $\alpha > 0$
and $\beta$ are two constants \cite{re:bates90}. Below $T_{ODT}$,
equilibrium structures of a wide variety of symmetries have been
predicted and experimentally observed \cite{re:fredrickson96}. Around
$f=0.5$ (symmetric mixture), a so called lamellar phase is observed,
in which nanometer sized layers of A and B rich regions alternate in
space.  When the copolymer is quenched from a high temperature to a
temperature $T < T_{ODT}$, a transient polycrystalline configuration results 
comprising
many lamellar domains (or grains), with lamellar normals of arbitrary
orientations.  In practice, full development of the equilibrium
state requires very long annealing times until substantial long
ranged order at the scale of the system size can be
achieved. However, the underlying ordering mechanisms and associated
rates that contribute to the large scale reorientation of the grains
are essentially unknown at present.  Not only partial ordering is
detrimental for some applications, but it can also lead to aging of
the material, as well as to potentially anomalous response to applied
stresses. We present below our analysis of one of the possible
mechanisms contributing to the kinetics of large scale reorientation of
lamellar domains, especially in connection
with the use of reciprocating shears to accelerate grain coarsening.

Imposing a reciprocating shear is one of the methods currently in use
to achieve long-ranged order of block copolymer microstructures. Early
work on the response of block copolymer blends to shears
\cite{re:hadziiaannou79,re:koppi92,re:koppi93} aimed at establishing
the dependence of $T_{ODT}$ on the shear rate, but the experiments
also revealed that the shear helped select specific lamellar orientations.
These observations have subsequently led to a large
number of groups attempting to quantify the type and degree of
ordering that can be achieved by reciprocating shears.  Early
experiments by Koppi et al. \cite{re:koppi92,re:koppi93} involved the
copolymer poly-(ethylene-propylene) - poly-(ethylethylene)
(PEP-PEE). Upon lowering the temperature below $T_{ODT}$, they
observed a transition to the so-called parallel lamellae at moderate
shear rates, but also an unexpected transition to perpendicular
lamellae at high frequencies (in parallel alignment, the layers are
normal to the shear gradient direction, whereas in perpendicular
alignment the layers are normal to the vorticity direction; see a
schematic representation in Fig. \ref{fig:alignment}). It was also
possible to induce the transition from parallel to perpendicular by
increasing the shear frequency, but this transformation was not
reversible. Koppi et al. interpreted the high frequency behavior as
shear disordering of the original configuration, followed by formation
of the perpendicular orientation. Instead, the parallel orientation
was argued to result from defect mediated growth.

This phenomenology is qualitatively consistent with the most complete
theoretical analysis to date due to Fredrickson
\cite{re:fredrickson94}, although one must note that his results were
obtained for steady shears instead. He used the same model equations
that we use below in our study, but explicitly allowed for thermal
fluctuations near $T_{ODT}$, as he was primarily interested in
modeling orientation selection at $T_{ODT}$ through anisotropic
fluctuation suppression by the shear flow. He approximated the effect
of the flow by introducing a modified order parameter mobility that
depended on the integrated flow over the volume of the sample. Given
the inverse characteristic decay time of concentration fluctuations,
$\dot{\gamma}^{*}$, he showed that for shear rates $\dot{\gamma} <
\dot{\gamma}^{*}$ the parallel orientation is preferred. In the
opposite limit, the first transition upon lowering the temperature
leads to a perpendicular orientation. Further decrease in temperature
led to transition back to a parallel structure. The range of existence
of the perpendicular orientation was argued to decrease with
increasing viscosity contrast between the two microphases.

However, the phenomenology just described is exactly reversed for a
poly-(styrene) - poly-(isoprene) (PS-PI) copolymer. Here the
perpendicular orientation is observed at low frequencies, and the
parallel orientation at large frequencies
\cite{re:patel95,re:gupta95}. There is disagreement in the results at
low frequencies, as Wiesner and coworkers find a parallel orientation
at low frequency \cite{re:maring97,re:leist99} in the same range
investigated by others.  The experimental results have been summarized
in \cite{re:tepe97}: for frequencies $\omega < \omega_{d}$ a parallel
orientation has been found in two out of four studies. At intermediate
frequencies $\omega_{d} < \omega < \omega_{c}^{\prime}$ the preferred
orientation is perpendicular, whereas for $\omega >
\omega_{c}^{\prime}$ the observed orientation is parallel. The
frequency $\omega_{d}$ is a characteristic inverse time of local
domain deformation, and $\omega_{c}^{\prime}$ is a frequency above
which chain relaxation dynamics dominates the storage modulus
$G^{\prime}(\omega)$.

A different line of theoretical investigation has shifted the focus of
study away from fluctuations at $T_{ODT}$ and into secondary
instabilities of a well developed lamellar pattern
\cite{re:kodama96,re:shiwa97,re:drolet99}.  Kodama and Doi
\cite{re:kodama96} used a cell dynamical model to study possible
instabilities of a lamellar pattern upon shearing. The numerical
results obtained motivated in turn an analytic stability analysis
that, now in the absence of flow, addressed lamellar stability against
a change in wavelength. Shiwa \cite{re:shiwa97} later investigated the
similarity between the amplitude or envelope equations describing slow
modulations of a lamellar structure and the same equations describing
roll distortions in Rayleigh-B\'enard convection. In the limit of
vanishing shear amplitude, he established the equivalence of the
stability diagram of both systems, and hence inferred the range of
stability of a lamellar phase against Eckhaus (longitudinal) and
zig-zag (transverse) instabilities. We were later able to obtain
lamellar solutions under uniform steady and oscillatory shears of
finite amplitude \cite{re:drolet99}, and to study their
stability. Stability boundaries of the base lamellar phase against
both melting and long wavelength perturbations were obtained for
transverse and parallel orientations. However, we did not address the
perpendicular orientation, nor the effect of viscosity contrast on the
stability diagram. The consideration of both a full three dimensional
geometry and viscosity contrast are the subject matters of this
paper.

We present a full three dimensional stability analysis of a uniform
lamellar phase under a reciprocating shear flow with an assumed linear
dependence of the shear viscosity on local monomer composition.  We
self-consistently determine the velocity field and monomer composition
and study the growth or decay of long wavelength perturbations of a
base lamellar phase. The stability analysis leads to a Floquet problem
for the perturbation amplitudes, which we solve numerically. First, we
find that viscosity contrast has a negligible effect on the stability
boundaries of the lamellar phase. Second, from the actual computation
of these boundaries we find that, in general, the region of stability
is larger for the perpendicular rather than the parallel
orientation. Third, the marginal mode for the transverse instability 
branch is typically
along the perpendicular orientation, so that the initial decay of both
parallel and transverse uniform states would lead to the appearance of
a perpendicular component.  We finally show that the orientation of
the marginal mode of instability can be generally understood from
geometrical considerations. At a zig-zag boundary, for example,
the marginal mode tends to be oriented along the direction that causes the 
largest decrease in lamellar wavelength by the shear.

\section{Mesoscopic model equations and lamellar phases}

At a mesoscopic level, a block copolymer melt is described by an order
parameter $\psi({\bf r})$ which represents the local density
difference between the two monomers constituting the copolymer.  The
corresponding free energy was derived by Leibler \cite{re:leibler80}
in the weak segregation limit (close to $T_{ODT}$), and later extended
by Ohta and Kawasaki \cite{re:ohta86} to the strong segregation
regime. If the temporal evolution of $\psi$ occurs through advection
by a flow field as well as through local dissipation driven by free
energy reduction, $\psi$ obeys a time dependent Ginzburg-Landau
equation \cite{re:oono88b,re:fredrickson94},
\begin{equation}
  \frac{\partial\psi}{\partial t} + ({\mathbf v}\cdot\nabla)\psi
   =M \nabla^2(- r \psi + u \psi^3 - K \nabla^2\psi) - M B \psi.
  \label{eq:psigoveq}
\end{equation}
with ${\bf v}$ the velocity, $B$ a parameter that characterizes the
amplitude of the long ranged interactions arising from the covalent
bond connecting the two sub-chains \cite{re:ohta86}, and $M$ a
phenomenological mobility or Onsager coefficient
\cite{re:fredrickson96}. The remaining parameters appearing in the
equation can be related to physical properties of the polymer chains
as follows \cite{re:kodama96}: $K = b^{2}/3$, where $b$ is Kuhn's
statistical length of the chain, $r = (2 \chi N - 7.2)/N$, and $B =
144/N^{2}b^{2}$.

Under the assumption of Newtonian behavior, the equation governing the
velocity field is an extended Navier-Stokes equation for an
incompressible fluid,
\begin{equation}
\label{eq:ns}
\rho_{0} \frac{\partial \mathbf{v}}{\partial t} + \rho_{0} 
({\mathbf v} \cdot {\mathbf \nabla})
{\mathbf v} = \nabla \cdot \left[ \eta \left( \nabla {\mathbf{v}} +
\nabla {\mathbf{v}}^{T} \right) \right]
- {\mathbf \nabla} p + \frac{\delta
{\cal F}}{\delta \psi}  {\mathbf \nabla} \psi , \quad \nabla
\cdot {\mathbf v} = 0,
\end{equation}
where $\rho_{0}$ is the (constant) density, $\eta$ the shear viscosity
of the fluid which may depend on $\psi$, $p$ is the pressure, and
appropriate boundary conditions for both $\psi$ and {\bf v} must be
introduced. The last term on the right-hand side of Eq. (\ref{eq:ns})
is required to ensure that there cannot be free energy reduction by
pure advection of $\psi$ \cite{re:gurtin96}. This term leads to the
creation of rotational flow by curved lamellae that is directed toward
their local center of curvature.

We first introduce dimensionless variables following
ref. \cite{re:drolet99}.  Lengths are scaled by $\sqrt{K/r}$ which is
proportional to the lamellar wavelength, time by $K/Mr^{2}$, the
characteristic monomer diffusion time, and order parameter $\psi$ by
$\sqrt{r/u}$. As described in that reference, only one dimensionless
group remains in Eq. (\ref{eq:psigoveq}), $B' = BK/r^{2}$. In
dimensionless units (we also omit the prime in $B$; it is assumed to
be a dimensionless coefficient in what follows),
Eq. (\ref{eq:psigoveq}) reads,
\begin{equation}
\label{eq:psieq}
\frac{\partial\psi}{\partial t} + ({\mathbf v}\cdot\nabla)\psi
   =\nabla^2(- \psi + \psi^3 - \nabla^2\psi) - B \psi.
\end{equation}
The mean field order-disorder transition occurs at $B_{c} =
1/4$ with a critical wavenumber $q_{c} = \sqrt{1/2}$. 

Equation (\ref{eq:ns}) can be further simplified by noting that the
order parameter diffusivity (proportional to of $Mr$) is much smaller
than the kinematic viscosity $\eta/\rho_{0}$, and that under typical
experimental conditions $\rho_{0} \omega d / \eta \ll 1$ as well,
where $\omega$ is the angular frequency of the oscillatory shear, and
$d$ the thickness of the block copolymer layer. We therefore adopt a
creeping flow approximation according to which the flow field
instantaneously relaxes to that determined by the instantaneous
configuration of the order parameter $\psi$ (and the no-slip boundary
conditions). We further neglect in our present study the term
$\frac{\delta {\cal F}}{\delta \psi} {\mathbf \nabla} \psi$ in the
right hand side of Eq. (\ref{eq:ns}). Therefore the only dependence of
this equation on the order parameter $\psi$ enters though the shear
viscosity $\eta = \eta(\psi)$, as we focus on the effect of viscosity
contrast between the microphases on the stability of a lamellar
structure.

An important restriction of our calculations is that the
fluid remains Newtonian. This is in line with previous calculations on
this system, but it is an approximation that needs to be removed in
future work, as already noted in the introduction. We consider here a
specific linear dependence of the shear viscosity on $\psi$,
\begin{equation}
\eta = \eta_0 + \eta_1\psi,
\label{eq:eta_def}
\end{equation}
where $\eta_{1}$ does not have to be small compared
to $\eta_{0}$. Within the approximation stated, and in dimensionless
variables, Eq. (\ref{eq:ns}) reduces to,
\begin{equation}
\label{eq:nsless}
0 = \nabla p + \nabla \left[ \left( 1 + \eta_{1} \psi \right) \left( \nabla
{\mathbf v} + \nabla {\mathbf v}^{T} \right) \right],
\end{equation}
where the dimensionless viscosity correction $\eta_{1}^{\prime} =
\sqrt{r/u} (\eta_{1}/\eta_{0})$ has been introduced (and the prime
removed).

The physical system under consideration is a layer of block copolymer,
unbounded in the $x_{1}$ and $x_{2}$ directions, and being uniformly
sheared along the $x_{3}$ direction (Fig. \ref{fig:alignment}). The
layer is confined between the stationary $x_{3}=0$ plane, and the
plane $x_{3}=d$ which is uniformly displaced parallel to itself with a
velocity ${\mathbf v}_{\mathrm{plane}} = \gamma \, d \, \omega \,
\cos(\omega t) \hat{\bf x}_{1}$, where $\hat{\bf x}_{1}$ is the unit
vector in the $x_{1}$ direction. $\gamma$ is the dimensionless strain
amplitude and $\omega$ is the angular frequency of the shear. In what
follows, this velocity is also expressed in the dimensionless
variables given above. In particular, the dimensionless wall velocity
can be written as ${\mathbf v}_{\mathrm{plane}}^{\prime} =
\gamma^{\prime} \cos (\omega^{\prime} t^{\prime})$ so that with the
scalings introduced earlier, $\gamma^{\prime} = \sqrt{2}
(\omega \tau_{D}) (q_{c}d) \gamma$, with $\tau_{D} = K/Mr^{2}$ the
(diffusion) time scale introduced earlier. Since for a realistic
system $q_{c} d \gg 1$, the relatively small values of
$\gamma^{\prime}$ required for instability (see below) implies that
our study addresses in practice shear frequencies much smaller than
the inverse diffusion time. Again, in what follows we restrict
ourselves to dimensionless variables and drop the primes.

We first summarize the results of ref. \cite{re:drolet99} in which the
viscosity was assumed uniform. We focused there on the weak
segregation limit $\epsilon = (B_{c} - B)/2B_{c} \ll 1$, in which the
solution for the monomer composition can be obtained perturbatively in
$\epsilon$,
\begin{equation}
\psi ({\mathbf r}) = 2 A(t) \cos ({\mathbf q} \cdot {\mathbf r}) +
A_{1}(t) \cos ( 3 {\mathbf q} \cdot {\mathbf r} ) + \ldots
\label{eq:sol2}
\end{equation}
where ${\mathbf r} = x_{1} \hat{\bf x}_{1} + x_{2} \hat{\bf x}_{2} +
x_{3} (\gamma \sin (\omega t) \hat{\bf x}_{1} + \hat{\bf x}_{3})$ so
that it can be thought of as having components in a non orthogonal basis 
set which follows the
imposed shear, and ${\mathbf q}=(q_{1}, q_{2}, q_{3})$ is the
wavevector in the corresponding reciprocal space basis set $\{
{\mathbf g}_{1} = \hat{\bf x}_{1} - \gamma \sin (\omega t) \hat{\bf
x}_{3}, {\mathbf g}_{2} = \hat{\bf x}_{2}, {\mathbf g}_{3} = \hat{\bf
x}_{3} \}$. Note that in this new coordinate system, perfectly ordered
configurations are stationary. Three orientations relative to the
shear can be defined as follows: $q_{3} \neq 0, q_{1}=q_{3} = 0$ is a
purely parallel orientation, $q_{2} \neq 0, q_{1}=q_{3}=0$ is a
perpendicular orientation, and $q_{1} \neq 0, q_{2}=q_{3}=0$ is a
transverse orientation.

In the absence of viscosity contrast, the velocity field is given by,
\begin{equation}
{\mathbf v}^0 = \gamma \omega \cos (\omega t) \; x_{3} \; \hat{\bf x}_{1}.
\label{eq:basev}
\end{equation}
Furthermore, by substituting Eq. (\ref{eq:sol2}) into
Eq. (\ref{eq:psieq}), the lowest order solution (${\cal O}
(\epsilon^{1/2})$) is given by
\begin{equation}
\psi ({\mathbf r}) = 2 A(t) \cos ({\mathbf q} \cdot {\mathbf r}) 
\label{eq:basepsi}
\end{equation}
where the amplitude $A(t)$ satisfies the equation \cite{re:drolet99},
\begin{equation}
\frac{dA}{dt} = \sigma[q^2(t)]A - 3 q^2(t)A^3,
\label{eq:ampl}
\end{equation}
with $q^2(t)=q_1^2 + [\gamma \sin (\omega t) q_1 -q_3]^2 + q_2^2$ and
$\sigma(q^2)=q^2-q^4-B$.  This equation can be integrated to give the
marginal stability boundaries, and the function $A(t)$ itself
\cite{re:drolet99}. From this analysis, a critical strain amplitude
$\gamma_{c}$, which depends on ${\bf q}$,
was identified such that for $\gamma < \gamma_{c}$ the
uniform lamellar structure oscillates with the imposed shear, but at
$\gamma > \gamma_{c}$ $A(t)$ decays to zero; i.e., the lamellar
structure melts, to use the terminology used in experiments.

The stability of the base lamellar pattern was then addressed in two
spatial dimensions by a Floquet analysis. The study was restricted to
the $(x_{1},x_{3})$ plane (i.e., $q_{2} = 0$), and therefore
restricted to transverse ($q_{3} = 0$) and parallel ($q_{1} = 0$)
lamellae only. Briefly, for sufficiently small $\gamma$ all
orientations were seen to retain a range of stability in wavenumber,
range that becomes narrower along the diagonal in the $(q_{1},q_{3})$
plane. The range of stability is further reduced with increasing
frequency. At moderate shear amplitudes (e.g., $\gamma = 0.4$), only
fully parallel or transverse lamellae remain stable, and there is also
a weak frequency dependence.  We extend below these results to three
spatial dimensions and to a fluid with a nonuniform shear viscosity,
function of the local monomer composition $\psi$. Our aim is to
incorporate into the stability analysis a different effective rheology
of the two microphases.

\section{Stability of a lamellar phase under shear}
\label{sec:lamellae}

We obtain in this section the flow in the melt that arises from the
nonuniform shear viscosity, and the resulting corrections to the base
lamellar solution. We then perform a self-consistent stability
analysis of the lamellar order parameter and flow field against long
wavelength perturbations.

There is now ample evidence that viscoelastic contrast between the
microphases affects orientation selection. We wish to investigate here
whether such a contrast substantially affects the stability of a
uniform lamellar phase to shear. This is an initial step in attempting
to understand the experimental phenomenology. For example, and as
briefly discussed in Section \ref{sec:introduction}, the qualitative
response to oscillatory shears in a system such as
poly-(ethylene-propylene) - poly-(ethylethylene) (PEP-PEE) is
qualitatively very different than that of, say, poly-(styrene) -
poly-(isoprene) (PS-PI). The former prefers the parallel orientation
at low frequencies and the perpendicular orientation at higher
frequencies, whereas the behavior is essentially reversed for the
latter. As emphasized by Fredrickson and Bates
\cite{re:fredrickson96}, the microphases of the PEP-PEE system are
well matched mechanically. However, PS is largely unentangled whereas
PI is entangled. Therefore a large contrast in the relaxation times of
the blocks is anticipated. Also recent experimental evidence by Winey
and coworkers \cite{re:polis99} suggests that the response of a
parallel configuration to shear can be qualitatively described by a
three region model. One central region in the vicinity of the covalent
bond between the A and B chains is relatively stiff, and responds
elastically under shear. This region is surrounded by two other
regions with a largely viscous response to the shear as the chains are
elongated. Before addressing the more general case of viscoelastic
contrast between the microphases, we study in this paper a fairly
simplistic situation in which the blend remains a Newtonian fluid, but
one in which the shear viscosity depends explicitly on the local
monomer composition $\eta=\eta(\psi)$, and hence the actual flow field
inside the melt depends on the order parameter configuration. We
follow a related study by Fredrickson \cite{re:fredrickson94} who
incorporated viscosity contrast by computing an averaged, effective
shear rate for a lamellar configuration, which then renormalized the
mobility coefficient in the time dependent Ginzburg-Landau equation
for the order parameter. We do consider here, however, the full
calculation of the flow field self-consistently, although for
practical convenience we restrict our analysis to the linear variation
given in Eq. (\ref{eq:eta_def}). In this case, we can solve for the
base flow field exactly, fact that considerably simplifies the
stability analysis.

We begin by writing Eq. (\ref{eq:nsless}) in components,
\begin{equation}
\partial_i p - \partial_j[(1 + \eta_{1} \psi) (\partial_i v_j + 
\partial_j v_i)] =0,
\label{eq:velgoveq}
\end{equation}
and consider periodic boundary conditions along the $\hat{\bf x}_{1}$
and $\hat{\bf x}_{2}$ directions. To facilitate the computation, we
replace no slip boundary conditions in the $\hat{\bf x}_{3}$ direction
by sheared periodic boundary conditions (i.e., periodic boundary
conditions in a frame of reference rigidly attached to the moving
plates. See, e.g., ref. \cite{re:drolet99} for further details).  We
then introduce the decomposition
\begin{equation}
\label{eq:vdec}
{\bf v} = {\bf v}^0 + {\bf u}.
\end{equation}
The governing equation for the velocity ${\bf u}$ is (with $\psi({\bf
r})=2A\cos({\bf q}\cdot{\bf r})$)
\begin{equation}
  -\p_ip_1+\p_j[(1 + \eta_{1} \psi)(\p_iv^0_j+\p_jv^0_i)]
         +\p_j[(1 + \eta_{1} \psi)(\p_iu_j+\p_ju_i)] = 0
\label{eq:vpert}
\end{equation}

In order to solve this equation, we introduce a new set of coordinates
$(x',y',z')$ such that $\hat{z'} = \hat{\bf q}$ (the $z'$ direction is
parallel to the {\em time dependent} wavevector of the lamellar
phase), and $\hat{y'}$ is perpendicular to the plane
$(q_z,0,q_x)$. Then $u_{y'}$ and $u_{z'}$ vanish in the lamellar
phase, and the equation for $u_{x'}$, after some algebra, is given by,
\begin{equation}
  \p_{z'}[(1 + 2\eta_{1} A(t) \cos qz')\p_{z'}u_{x'}] 
              = \gamma\eta_1 A(t) q\beta\sin qz',
  \label{eq:vel}
\end{equation}
with $\beta=2[-(q_x^2-q_z^2)^2-q_y^2(q_x^2+q_z^2)]/qn_x$ and $n_x =
|(q_z(q_x^2-q_y^2-q_z^2), 2q_xq_yq_z,$ $-q_x(q_x^2+q_y^2-q_z^2))|$,
where the wavevector ${\mathbf q}$ is a function of time as the orientation of
the base lamellar structure adiabatically follows the imposed shear flow.
Equation (\ref{eq:vel}) can now be integrated twice to obtain $u_{x'}$.
It is important to note that for a purely perpendicular base state $\beta = 0$,
and hence the flow correction ${\bf u} = 0$ in this case. This is in 
agreement with the results of Fredrickson \cite{re:fredrickson94}.

A general property of the solution for the flow perturbation ${\mathbf
u}$ that follows from the linear form of the viscosity contrast is
that the velocity is always perpendicular to ${\bf q}$, i.e., parallel
to the lamellar planes which are the planes of constant $\psi$.
Therefore the advection term ${\mathbf u} \cdot \nabla \psi = 0$ for a
lamellar phase, and hence the base field $\psi$ is unaffected by the
flow.  Note that this result is general as long as the order parameter
is a function of one spatial direction only. In the more general
cases, only the functional dependence in $\cos (z')$ and $\sin (z')$
in Eq.~(\ref{eq:vel}) would be different. In short,
Eq. (\ref{eq:basepsi}) remains the solution for the order parameter
when the viscosity contrast is given by Eq.~(\ref{eq:eta_def}). The base flow
field can be obtained from the solution of Eq.~(\ref{eq:vel}) and
Eq.~(\ref{eq:vdec}).

We next address the stability of the solution Eqs. (\ref{eq:basev})
and (\ref{eq:basepsi}) against long wavelength perturbations at fixed
$\epsilon$ \cite{fo:peilong4_1}. We consider solutions of the form
\begin{eqnarray*}
  \psi &=& \psi_1 + \psi_2 \\
  {\bf v} &=& {\bf v}^0 + {\bf u} + {\bf w}
\end{eqnarray*}
where $\psi_{1}$ is the base solution given in Eq. (\ref{eq:basepsi}),
${\mathbf u}$ the solution of Eq. (\ref{eq:vel}), and,
\begin{equation}
   \psi_2 = \psi_{2+} \; e^{i\kr+i\Kr} + \psi_{2-} \; e^{i\kr-i\Kr} + \cc
   \label{eq:purb}
\end{equation}
If the wavevector of the perturbation ${\bf Q}$ is parallel to the
base wavevector ${\bf q}$, the perturbation is said to be of the
Eckhaus type. In the case of ${\bf Q} \perp {\bf q}$, we are
considering a zig-zag perturbation. Due to our choice of shear
periodic boundary conditions, and unlike more standard analyses of
long wavelength instabilities, the perturbation wavevector ${\bf Q}$
is time dependent and periodic with the same periodicity as the shear
(the same is true for ${\bf q}$).

The stability of periodic solutions of the current model in the
absence of flow has been extensively studied
\cite{re:cross93,re:shiwa97}. There is a range of wavenumbers $q$
within which the lamellar states are stable. This range of stability
increases with the distance to threshold $\epsilon$. At large
wavenumbers, lamellae undergo an Eckhaus instability which tends to
lowers the value of $q$ by eliminating lamellar layers. On the other
hand, at small wavenumbers the wavenumber is increased via a zig-zag
instability.  As we show below, the imposed oscillatory shear has
important consequences for these two instabilities, especially for the
zig-zag case. In particular, we show that the way in which a zig-zag
instability leads to a readjustment of the lamellar wavelength depends
strongly on the relative orientation between the base lamellae and the
shear direction.

In order to derive an evolution equation for the two amplitudes
$\psi_{2+}$ and $\psi_{2-}$ we need the velocity field ${\bf
w}$. Since the convective term in Eq. (\ref{eq:psigoveq}) is ${\bf
v}\cdot\nabla\psi = ({\bf u} + {\bf w})\cdot \nabla(\psi_1 + \psi_2)$,
we only need to retain Fourier components of the form $e^{\pm i\kr\pm
i\Kr}$. Therefore one only needs the Fourier components $e^{\pm
2i\kr}$ for ${\bf u}$ and $e^{\pm 2i\kr\pm i\Kr}$ for ${\bf w}$. The
governing equation for the second order velocity ${\bf w}$ reads,
\begin{eqnarray*}
  -\p_ip_2 + \p_j[(1 +\eta_1\psi_1)(\p_iw_j + \p_jw_i)] \hskip 4cm & & \\
        ~  + \p_j[\eta_1\psi_2(\p_iu_j+\p_ju_i)] 
          + \p_j[\eta_1\psi_2(\p_iv^0_j+\p_jv^0_i)] &=& 0.
\end{eqnarray*}
This equation is solved by transforming again to the $(x',y',z')$
coordinate system. The sum of the third and fourth terms is the
inhomogeneous term in the equations and is of the form of ${\bf
h}(qz')e^{i\Krp}+\cc$. Therefore we write the second order velocity as
\begin{eqnarray*}
  {\bf w} &=& e^{i\Krp}{\bf \overline{w}}(qz') + \cc \\
  p_2 &=& e^{i\Krp} \overline{p}_2(qz') + \cc
\end{eqnarray*}
which leads to the equations
\begin{eqnarray}
  -\p_i \overline{p}_2(qz') + (iQ'_j+\p_j)[1 + \eta_{1} \psi_1(qz')] \times 
           \qquad & & \cr
     [(iQ'_i+\p_i) \overline{w}_j(qz')+(iQ'_j+\p_j)
     \overline{w}_i(qz')] &=& h_i(qz').
\label{eq:wsol}     
\end{eqnarray}
We now solve for the functions ${\bf \overline{w}}(qz')$ and
$\overline{p}_2(qz')$ numerically together with the incompressibility
condition,
\begin{equation}
iQ_{x'}\overline{w}_{x'} + iQ_{y'}\overline{w}_{y'} + 
(iQ_{z'}+\p_{z'})\overline{w}_{z'} = 0.
\label{eq:incomp}
\end{equation}

Since the velocity ${\bf w}$ is proportional to $\psi_2$, and given
the latter's decomposition in its components $e^{2i\kr}$ and
$^{i\Kr}$, we find it useful to express ${\bf w}$ as
$$
{\bf w} = \left( 
             {\bf w}^{++}e^{2i\kr}\psi_{2+} 
           + {\bf w}^{+-}e^{-2i\kr}\psi_{2+}
           + {\bf w}^{-+}e^{2i\kr}\psi_{2-}^* 
           + {\bf w}^{--}e^{-2i\kr}\psi_{2-}^* \right) e^{i\Kr} + \cc
$$
Also by considering the Fourier expansion of ${\bf u}$,
$$
{\bf u} = {\bf u}_1 e^{i\kr} + {\bf u}_2 e^{2i\kr} + \cdots + \cc
$$
we arrive at the (linearized) system of equations obeyed by the perturbation
$\psi_{2}$,
\begin{equation}
  \frac{\partial}{\partial t}\left[
    \begin{array}{c} \psi_{2+} \\ \psi_{2-}^* \end{array} \right]
    = \left[
      \begin{array}{cc}
         H_{11}(t) & H_{12}(t) \\ H_{21}(t) & H_{22}(t)
      \end{array}
      \right]
      \left[
    \begin{array}{c} \psi_{2+} \\ \psi_{2-}^* \end{array} \right ] .
  \label{eq:floquet}
\end{equation}
This is the central result of this section. The
matrix elements are given by,
\begin{eqnarray*}
  H_{11}(t) &=& -l_+-l_+^2+B+6A(t)^2l_+ + iA(t){\bf w}^{++}\cdot{\bf q} \\
  H_{12}(t) &=& 3A(t)^2l_+ + iA(t){\bf w}^{-+}\cdot{\bf q}
                 + i {\bf u}_2\cdot({\bf q}-{\bf Q}) \\
  H_{12}(t) &=& 3A(t)^2l_- - iA(t){\bf w}^{+-}\cdot{\bf q}
                 - i {\bf u}_2\cdot({\bf q}+{\bf Q}) \\
  H_{22}(t) &=& -l_--l_-^2+B+6A(t)^2l_- - iA(t){\bf w}^{--}\cdot{\bf q}
\label{eq:matrix}
\end{eqnarray*}
with $l_{\pm}=-({\bf q}+{\bf Q})^2$. First we note that the matrix
elements are periodic in time, with the same periodicity as the
imposed shear. Therefore the stability analysis constitutes a Floquet
problem for the amplitudes $\psi_{2+}$ and $\psi_{2-}$ which we solve
numerically. Briefly, the function $A(t)$ is evaluated numerically for
a given lamellar orientation and shear flow parameters. The flow
correction due to viscosity contrast ${\bf u}$ in the base state is
known analytically, but the flow perturbation ${\bf w}$ is obtained
numerically by solving Eqs. (\ref{eq:wsol}) and (\ref{eq:incomp}).
Once the matrix elements in Eq.  (\ref{eq:matrix}) have been
numerically evaluated, the Floquet problem is recast as an eigenvalue
problem, which is then solved numerically.  For fixed $\epsilon,
\gamma, \omega$ and $\eta_{1}$ the stability boundaries are determined
as the loci of ${\bf q}$ at which the eigenvalue function $\sigma(
{\bf Q})$ changes from a maximum to a saddle point at ${\bf Q} = 0$.

We finally note that the fact that the Floquet problem only involves
the amplitudes $\psi_{2+}$ and $\psi_{2-}$ and not ${\bf w}^{++}, 
{\bf w}^{+-}, {\bf
w}^{-+}$ and ${\bf w}^{--}$ is a direct consequence of the creeping
flow approximation introduced: the velocity field is slaved to the
composition field.

\section{Results and discussion}
\label{sec:results}

\begin{figure}
\putgraph{15cm}{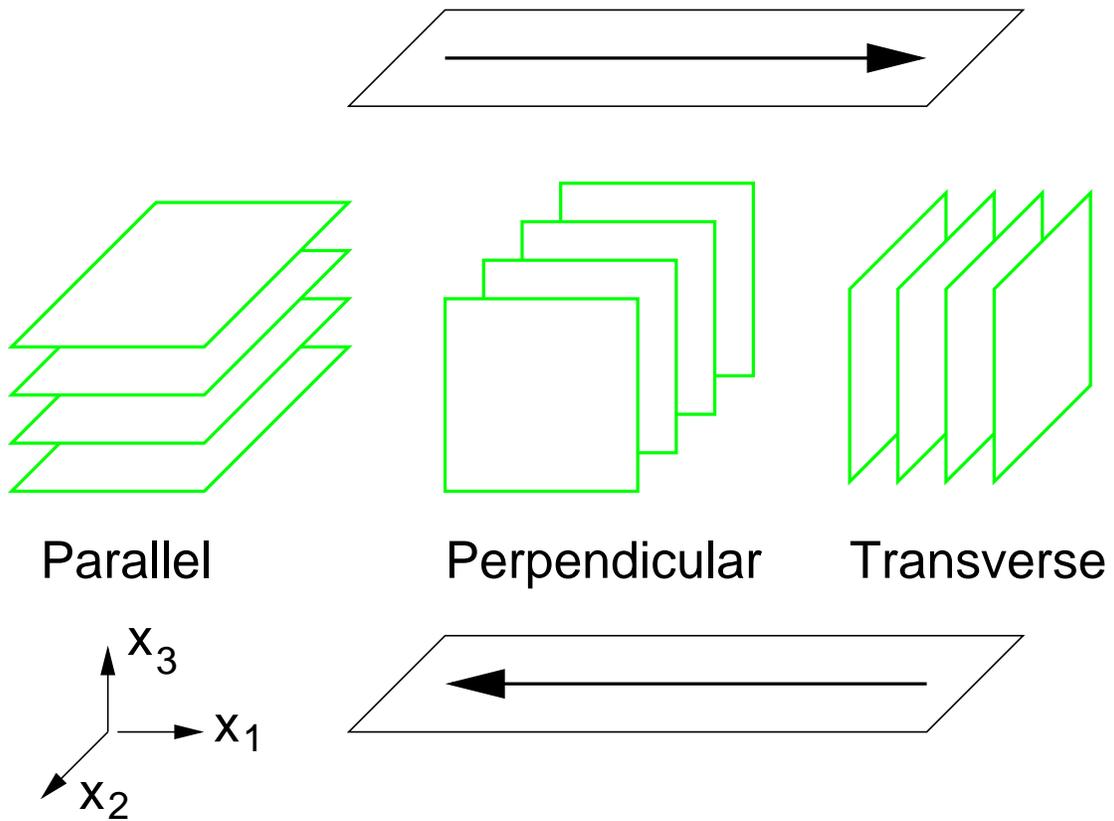}
\caption{Schematic representation of the geometry considered including
the shear direction, and the three different lamellar orientations discussed
in the text.}
\label{fig:alignment}
\end{figure}

As illustrated in Fig.~\ref{fig:alignment} and discussed in Section
\ref{sec:introduction}, the configuration considered is an unbounded
layer being sheared along the $x_1$ direction, and with a velocity
gradient in the $x_3$ direction (and hence the vorticity vector is
along the $x_2$ direction). Three particular orientations of the
lamellae are of special concern in relation with the issue of
orientation selection under shear: parallel, perpendicular, and
transverse, as indicated schematically in Fig.~\ref{fig:alignment}.
These three special cases are to be discussed first, although we later
present the results of the stability analysis for an arbitrary
orientation of the lamellae.

We again note that we do not focus on the formation of a particular
orientation from an initially disordered configuration, but rather on
the simpler problem of ascertaining the region of stability of a base
lamellar state of a given orientation as a function of the shear
rate. Of course, linearly unstable orientations are not expected to be
observable in experiments, not even locally during the transient
evolution of polycrystalline configurations. Our analysis, however,
does not address the determination of the basin of attraction of each
orientation from an initially disordered configuration, and hence
cannot completely answer the orientation selection problem. However,
as we will show below, the regions of stability against long
wavelength perturbations are quite small, and therefore our results do
provide some guidance for the orientation selection problem.

One of the main conclusions from our numerical study is that velocity
field corrections due to the assumed viscosity contrast between the
two microphases phases have a negligible effects on those stability
boundaries that we consider. We believe that this is due to two
reasons. First, we have shown that viscosity contrast of the assumed
form has no effect on the base state of perfectly parallel lamellae
(c.f. Eq.~(\ref{eq:vel}) and the paragraph that follows). Second, both
Eckhaus and zigzag instabilities are long wavelength instabilities.
Consider, for example, a transverse zig-zag perturbation of a parallel
state. Since the induced flow ${\bf u}$ alternates in direction on 
consecutive parallel planes, it will distort the zig-zag perturbation, but 
the effect will be very small for a long wavelength perturbation.
Hence in what follows we concentrate the discussion on the effects
that follow from lamellae orientation, keeping in mind that they are
quite independent of the value of $\eta_{1}$ considered.

The stability of a particular lamellar phase (with a base wavevector
${\bf q}$) is determined by the growth or decay of perturbations of
wavevector ${\bf q}+{\bf Q}$ which is in turn given and by the sign of
the corresponding eigenvalues of the Floquet problem defined by
Eq. (\ref{eq:floquet}). Also, the orientation of the marginally
unstable mode ${\bf Q}$ can be used as an indicator of the orientation
of the emerging structure. Of course, in an extended sample in which
multiple orientations coexist, possibly with a subset of them becoming
linearly unstable, it will be their nonlinear competition that will
determine the asymptotically selected orientation. Nevertheless, we
believe that it is still useful to catalog the orientation of the
marginal mode in the case of the three basic orientations of the
lamellae: parallel, perpendicular and transverse.

Before we turn to the numerical results of the Floquet analysis, we also
mention that we find it useful to interpret them in terms of a simple
geometrical construction of how the imposed shear affects the lamellar
wavelength. If one were to neglect any flow corrections due to
viscosity contrast, and if the lamellar base state were to follow the
base shear flow adiabatically, then a lamellar phase with initial
wavevector ${\bf q}(t=0)=(q_1,q_2,q_2)$ in the Cartesian or laboratory
frame of reference, evolves according to
\begin{equation}
{\bf q}(t)= (q_1, q_2, - \tilde{\gamma} q_1 + q_3),
\label{eq:wavevec}
\end{equation}
with $q_{1}, q_{2}$ and $q_{3}$ constant and $\tilde{\gamma} = \gamma
\sin (\omega t)$. We will see that in most cases the marginal mode at the
zig-zag instability is that which leads to the largest increase in wavenumber
by the shear.

\begin{figure}
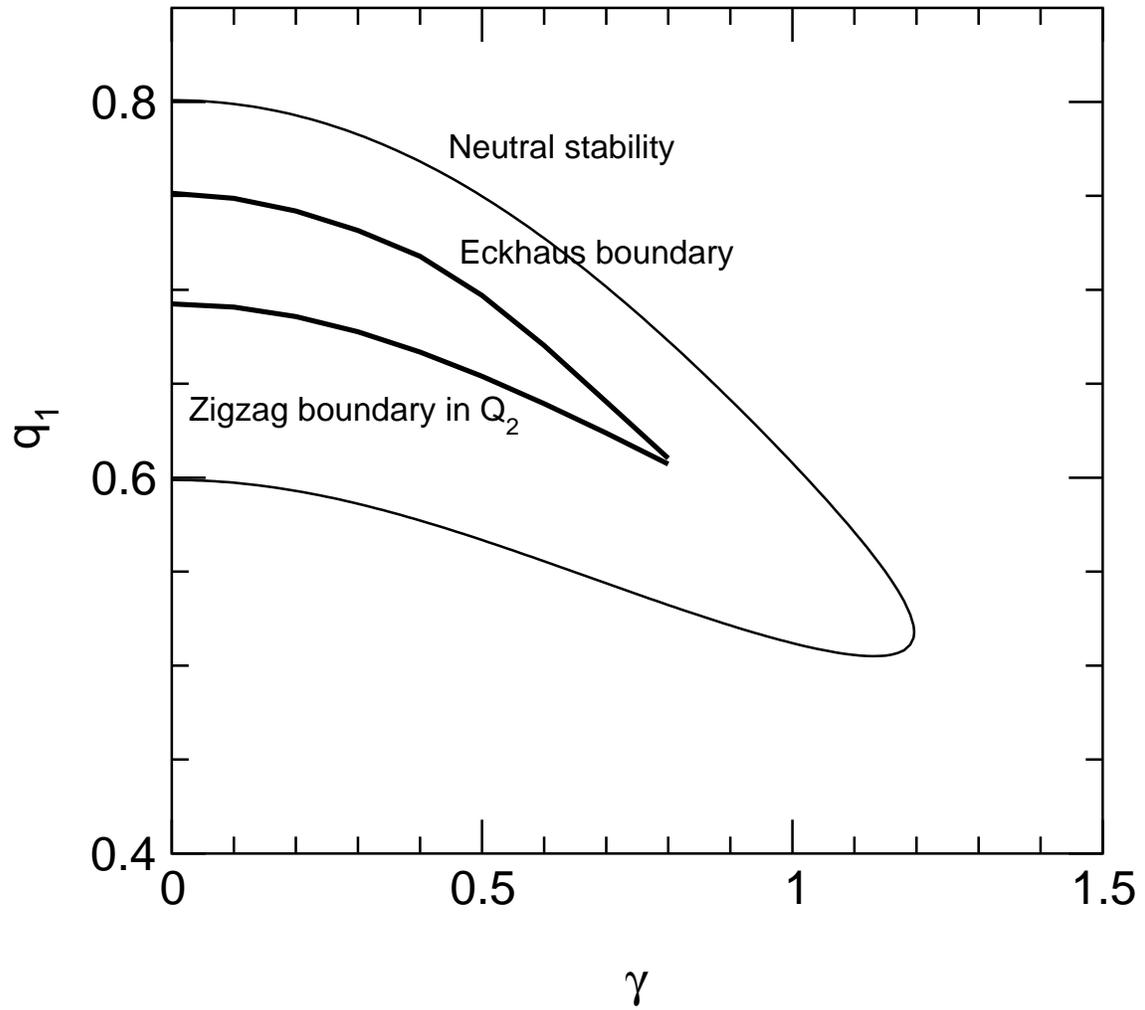

\putgraph{15cm}{transverse.eps}
\caption{Neutral stability curve and secondary instability boundaries
for an initially transverse lamella with
 $\omega=0.01$ and $\epsilon=0.04$.}
\label{fig:transverse}
\end{figure}

First consider initially transverse lamellae, i.e., with ${\bf q}$
along the $x_1$ direction. From Fig.~\ref{fig:alignment} it is clear
that the shear flow will tilt the layers, and in doing so decrease
their wavelength.  The stability diagram for this case has been obtained by
numerical solution of the Floquet problem for $\omega=0.01$ and
$\epsilon=0.04$ and is shown in Fig.~\ref{fig:transverse} as a function of
the shear amplitude $\gamma$. The outside bounding curves are the
neutral stability boundaries so that only within this range a
nonlinear solution for $\psi$ exists, with an amplitude $A(t)$ given
by Eq.~(\ref{eq:ampl}). The bending of the curves toward smaller
values of $q$ as $\gamma$ increases can be qualitatively understood by
noting that the oscillatory shear leads to a decrease in the lamellar
wavelength. Thus at higher values of $\gamma$, the region of existence
of finite amplitude solutions shifts toward larger wavelengths to
compensate for a larger reduction in wavelength by the flow.  At large
enough $\gamma$, however, nonlinear solutions cease to exist.

Only nonlinear solutions within the inner region in
Fig.~\ref{fig:transverse} are stable against long wavelength secondary
instabilities.  Both Eckhaus and zig-zag type instabilities occur as
shown in the figure. The Eckhaus instability is a longitudinal phase
instability (${\bf Q}\parallel{\bf q}$) so that through coupling with
the amplitude it leads to an amplitude modulation in the same
direction as the base periodicity. In general, this instability
appears in the large wavenumber range of the diagram and is
qualitatively interpreted as an instability that leads to a decrease
in the wavenumber by eliminating lamellar layers through the amplitude
modulation. On the other hand, in the range of small $q$, a zig-zag
instability with ${\bf Q}\perp{\bf q}$ can lead to an increase in $q$
to return the unstable lamella to the stable range. In the absence of
shear, perturbations to a base state defined by ${\bf q}=q\hat{\bf x}_1$
along $Q\hat{\bf x}_2$ or $Q\hat{\bf x}_3$ are equivalent.  Under
shear, however, the degeneracy is broken, and we find from the
numerical Floquet analysis that the marginal mode for the zig-zag
instability is $Q\hat{\bf x}_2$, i.e., along the perpendicular
direction.
In summary, the large wavenumber instability is of the Eckhaus type
and does not lead to lamellar reorientation. The small wavenumber
instability is of zig-zag type, and leads to a growing component along
the perpendicular direction.

We next turn to a geometrical interpretation of these results based on a rigid
distortion of the lamellar structure as discussed above. The marginal
wavevector for an unstable zigzag perturbation can be either ${\bf q}
+ {\bf Q} = (q,Q,0)$ which leads to growth of a perpendicular
component, or ${\bf q} + {\bf Q} = (q,0,Q)$ that leads to a parallel
orientation. The effective wavenumber of the distortion in each case
is given by
\begin{eqnarray*}
\hbox{perpendicular zigzag} \quad |{\mathbf q} + {\mathbf Q} |^{2}(t) & = &
 |(q,Q,\tilde{\gamma} q)|^2 =(1+\tilde{\gamma}^2)q^2+Q^2 
\label{eq:transziga} 
\\
\hbox{parallel zigzag} \quad |{\mathbf q} + {\mathbf Q} |^{2}(t) & = & 
 |(q,0,\tilde{\gamma} q+Q)|^2
      =(1+\tilde{\gamma}^2)q^2+2\tilde{\gamma} q Q +Q^2
\label{eq:transzigb}.
\end{eqnarray*}
Under oscillatory shear $\tilde{\gamma}$ changes sign and both cases
lead to a wavenumber increase, although given the sign change, it is
not possible to tell which mode is more effective at increasing the
wavenumber following the instability. This is the only case in which
this geometric interpretation does not lead to a selected marginal
mode consistent with the Floquet analysis. We note, however,
that if the shear
were stationary instead $\tilde{\gamma}$ increases monotonically, and
a zig-zag instability along the parallel direction would provide for a
larger increase in wavenumber.

\begin{figure}[t]
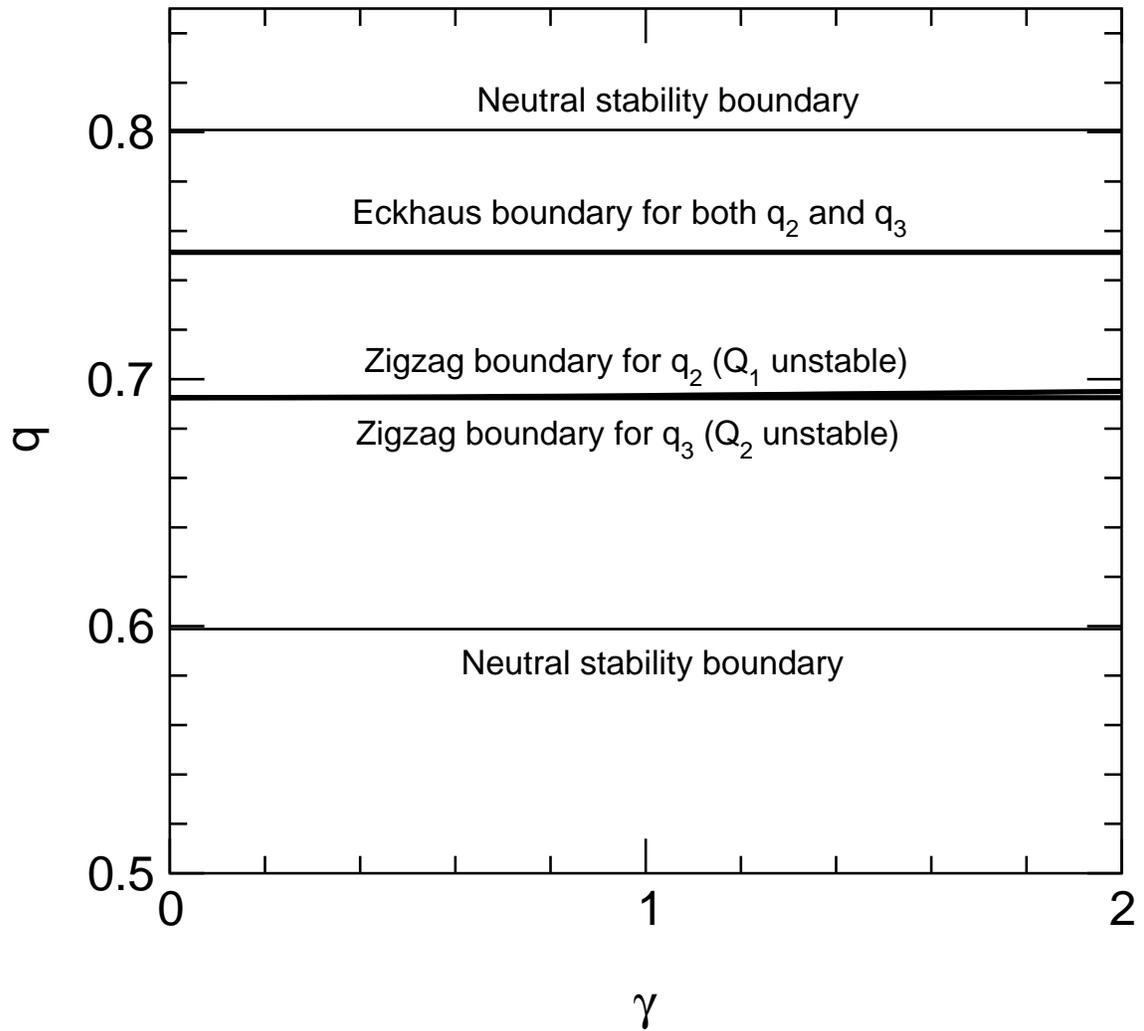

\putgraph{15cm}{parallel.eps}
\caption{Neutral stability curve and secondary instability boundaries
for initially parallel and perpendicular lamellae with
$\omega=0.01$ and $\epsilon=0.04$.}
\label{fig:parallel}
\end{figure}

We now consider a base lamellar phase oriented parallel to the shear.
Figure~\ref{fig:parallel} shows the stability boundaries computed
numerically.  As seen in the figure, secondary instabilities are again
of the Eckhaus type at large $q$, and of zig-zag type at small $q$. In
this case the location of the boundary is nearly independent of
$\gamma$, i.e., the shear flow has no effect on the Eckhaus
instability. This can be understood geometrically from
Fig.~\ref{fig:alignment} since the parallel orientation is unaffected
by the shear, and an Eckhaus instability of a parallel orientation
would also lead to a growing mode along the parallel orientation. The
zig-zag boundary at small wavenumbers also remains constant and
unaffected by the shear.  Nevertheless shear flow does introduce a
distinction between the two possible zig-zag modes along the
perpendicular and transverse directions respectively.  The former,
given by ${\bf q}=(0,Q,q)$, will not be advected by the shear according
to the Eq.~(\ref{eq:wavevec}). The latter is given by ${\bf
q}=(Q,0,q)$, and it has its wavenumber advected as $q(t) =
(q-\tilde{\gamma} Q)^2 + Q^2$. Therefore the shear leads to a net
wavenumber increase only when the instability is along the
transverse direction. The results of the Floquet analysis are
consistent in this case with the orientation that produces the largest
wavenumber increase.

We finally discuss the case of initially perpendicular lamellae.  The
stability boundaries for this case are also shown in
Fig.~\ref{fig:parallel}. Again, the Eckhaus boundary is unaffected by
the shear. This can be understood from Fig.~\ref{fig:alignment} since
the perpendicular orientation does not couple to the flow. An Eckhaus
mode is along the perpendicular direction as well, and hence remains
unaffected by the shear.  On the other hand there is a very weak
dependence of the zig-zag stability boundary on the shear amplitude,
and we find that the marginal mode is along the transverse
direction. This is also consistent with the geometric interpretation
given earlier since an imposed shear flow will always increase the
wavenumber for a transverse zig-zag mode $(Q,q,0)$, but has no effect
on a parallel zig-zag mode $(0,q,Q)$:
\begin{eqnarray*}
\hbox{transverse zigzag} \quad | {\mathbf q} + {\mathbf Q} |^{2} (t) &
= & |(Q,q,\gamma Q)|^2 = q^2+(1+\gamma^2)Q^2 \\ \hbox{parallel zigzag}
\quad | {\mathbf q} + {\mathbf Q} |^{2} (t) & = & |(0,q,Q)|^2 =
q^2+Q^2
\end{eqnarray*} 

Our results for the the secondary instabilities of the three
particular lamellar orientations can be classified according to the
orientation of the marginal mode as follows:
$$
\begin{array}{lcl}
  \hbox{Transverse lamellae} & \rightarrow & 
            \hbox{zig-zag to perpendicular orientation} \\ 
  \hbox{Parallel lamellae} & \rightarrow & 
            \hbox{zig-zag to perpendicular orientation} \\ 
  \hbox{Perpendicular lamellae} & \rightarrow & 
            \hbox{zig-zag to transverse orientation}
\end{array}
$$

If one assumes that the orientation of the marginal mode sets a
preference for the final orientation of the emerging stable lamellar
structure, our results suggest that the perpendicular alignment is the
preferred orientation under the action of the oscillatory shear,
independently of viscosity contrast (within the Newtonian viscosity model
adopted).  Even in the case in which an instability of a perpendicular
orientation leads to a transverse state, a shear flow of small
amplitude would drive it back to the perpendicular orientation,
presumably now with a wavenumber inside the stable region. We however
need to be remain cautious about the generality of this conclusion as
a complex dynamical evolution could follow the initial decay after a
zig-zag instability. For example, our earlier numerical results in
ref. \cite{re:drolet99} showed that a zig-zag instability may lead to
kink-band formation, and yet to another orientation change as the
bands themselves become unstable.

We finally present the stability diagram in full three dimensional
${\bf q}$ space for a base lamellar state of arbitrary orientation
relative to the shear. In the absence of flow, all lamellae with
wavenumbers near the marginal value $q_c$ are stable.  Hence the
stability region is a spherical shell in ${\bf q}$ space. Under shear,
any lamella with a wavevector that has a significant component along
the transverse direction becomes unstable, and the region of stability
becomes compressed along the $q_1$ direction.  For sufficiently large
$\gamma$, the stable region adopts a toroidal shape around the
$q_2$--$q_3$ plane with a small projection
in the $q_1$ direction that depends on $\gamma$. Although we have
argued earlier that lamellae with a too small wavenumber will become
unstable to a zig-zag mode and evolve into a new state oriented along
the perpendicular direction, this by no means implies that a stable
parallel orientation cannot exist (although perhaps within a narrow
range of wavenumbers).

\begin{figure}[t]
\putgraph{15cm}{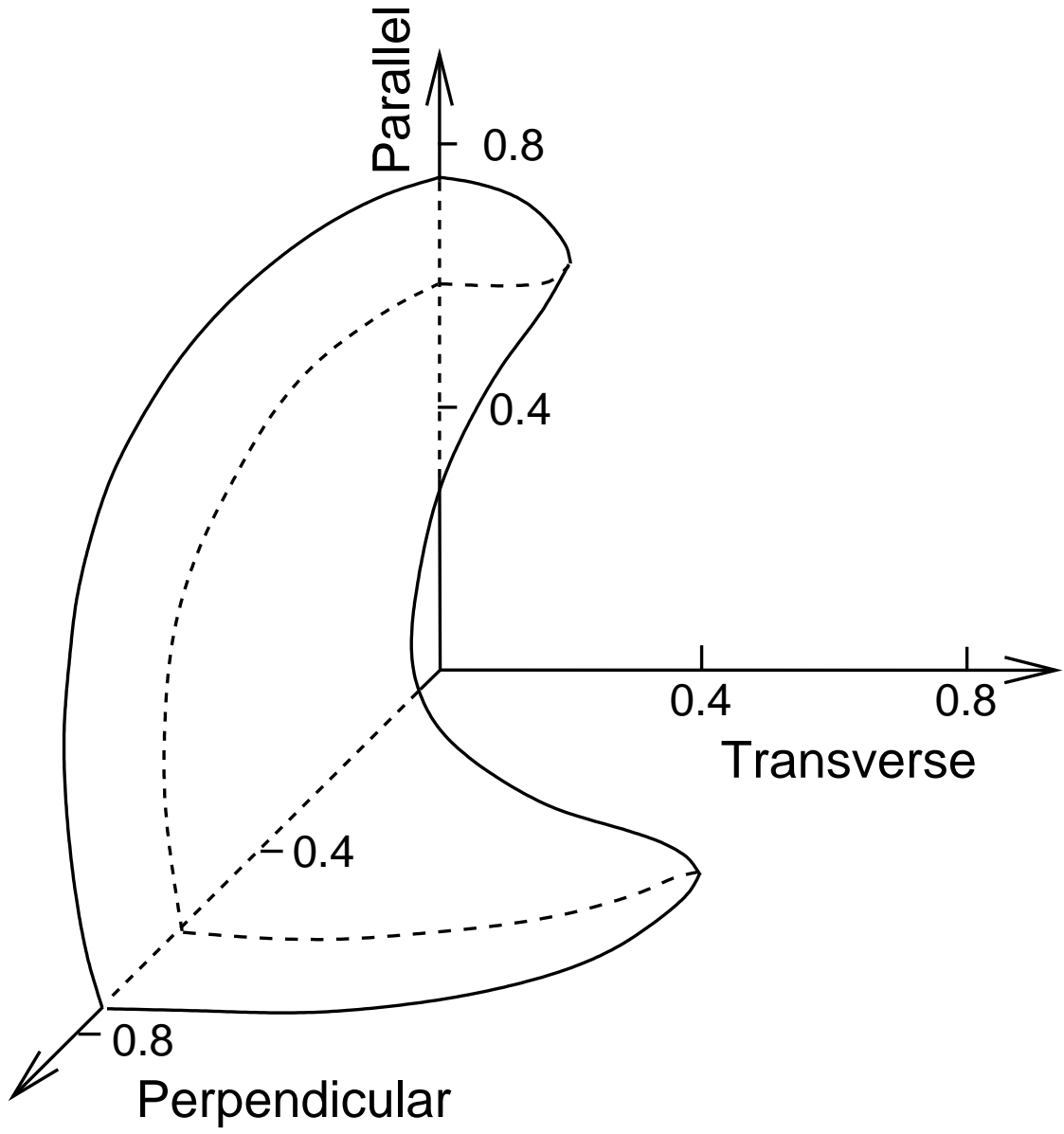}
\caption{Region of stability of a uniform lamella as a function of its
orientation. The axes correspond to transverse ($q_{1}$), 
perpendicular ($q_{2}$), and parallel ($q_{3}$) orientations. The 
surface has been determined by numerical solution of the Floquet problem
defined by Eq. (\ref{eq:floquet}) with $\omega=0.01$, $\gamma=1$, and 
$\epsilon=0.04$}
\label{fig:qtotal}
\end{figure}

As an example, Fig.~\ref{fig:q23reg} shows the complete stability boundaries
along both $q_2$--$q_1$ and $q_3$--$q_1$ planes for $\epsilon=0.04$
and $\gamma=1$. Note how the stable region in the $q_2$--$q_1$ plane is
significantly larger than in the $q_3$--$q_1$ plane. This is also evidenced
in Fig.~\ref{fig:qtotal}
by the asymmetry in the toroidal stability region which is wide
near the $q_2$ axis (the perpendicular direction)
but narrow near the $q_2$ direction (the parallel axis). The
implication of this result is that if an initially disordered state is
comprised of many domains locally oriented along arbitrary directions,
one would expect that a larger portion of the sample would remain in
the perpendicular state, as it corresponds to the one with the largest
region of stability in ${\bf q}$ space. This argument is distinct from
that given above involving the orientation of the marginal model
following a secondary instability of a perfectly ordered lamellar
structure, but it suggests the same dominant orientation.

In addition to the detailed numerical computation of the stability
region in ${\bf q}$ space, the relative sizes of the stability regions
for parallel and perpendicular orientations may be estimated from the
geometric interpretation given above. Consider first an almost
perpendicular state with a wavevector of a magnitude $q_0$ with a
small component $\Delta q$ in the transverse direction. Under the
action of the shear, Eq.~(\ref{eq:wavevec}) gives for the time
dependent wavenumber
$$
q(t) = \left[ \Delta q^2 + q_0^2 + (\tilde{\gamma}\Delta q)^2\right]^{\frac12}
    \approx q_0\left[ 1 + \frac12(1+\tilde{\gamma}^2)\left(\frac{\Delta q}{q_0}
             \right)^2 \right] .
$$
Therefore the wavenumber increases with the shear as $(\Delta q/q_0)^2$. On 
the other hand, an almost parallel state has a wavenumber
$$
q(t) = \left[ \Delta q^2 + (\tilde{\gamma}\Delta q + q_0)^2\right]^{\frac12}
    \approx q_0\left( 1 + \tilde{\gamma}\frac{\Delta q}{q_0} \right),
$$
and therefore it increases only linearly with $\Delta q/q_0$. These
two relations essentially yield the dependence of the width of the
toroidal region in Fig.~\ref{fig:qtotal} along the transverse
direction, and therefore the fact that the relative extent of the
perpendicular region of stability is larger than the parallel
region.

Insofar our results could be used to infer the dominant orientation
under shear, our conclusions would differ from those of Fredrickson 
\cite{re:fredrickson94}, from low frequency experimental evidence 
in PEP-PEE, but agree with some low frequency experimental 
evidence in PS-PI. One must caution, however, that as discussed in ref. 
\cite{re:fredrickson94}, it is not clear
which of the two systems PEP-PEE or PS-PI has a larger value of $\eta_{1}$,
or how close they are to the assumed Newtonian behavior. Therefore we must
consider our results to be a baseline against which to compare future work 
involving true viscoelastic contrast between the microphases.

Finally it is interesting to address a possible reason for the
qualitative discrepancy between the conclusions of our work and those
of Fredrickson's concerning the effect of viscosity contrast. Although
he discussed the case of steady shear, our results are confined to low
frequencies, and are only weakly dependent on frequency. Therefore the
discrepancy in conclusions is not likely to arise from the time
dependence of our solutions. Briefly, he found a transition from a
high temperature disordered state to parallel lamellae at low shear
rates, and to perpendicular lamellar at high shear rates. In the
latter case, and upon further decrease in temperature, he predicted
another transition to a parallel state. The location of this second
transition line depended on the viscosity contrast between the
microphases. Instead, we see no discernible effect arising from
viscosity contrast. A possible qualitative explanation can be given as
follows: Our calculation is conducted at an externally imposed shear
rate $D = v_{p}/d$ in dimensional units (here and in what follows),
where $v_{p}$ is the velocity of the solid boundary. Consider a
parallel configuration comprised of only two planar layers with
uniform shear viscosity $\eta_{0}+\eta_{1}$ and $\eta_{0} - \eta_{1}$
respectively. The average shear rate of this configuration is given by
$$
\langle D \rangle = \frac{1}{d} \left[ \frac{d}{2} 
\frac{(v'-0)}{d/2} + \frac{d}{2} \frac{(v_{p}-v')}{d/2} \right] = D,
$$
where $v'$ is the (unknown) speed at the boundary between the two layers.
In this case, the average shear rate is independent of the viscosity contrast.
On the other hand and following Fredrickson, if the boundary conditions 
involve an imposed shear
stress $\sigma$, the conclusion is different. The shear rate for the
base state is now $D = \sigma / \eta_{0}$, the same as in the previous case.
However, in the creeping flow approximation considered, the shear stress
is uniform in the fluid, but the shear rate is different in the two
layers,
$$
D_{1} (\eta_{0}-\eta_{1}) = D_{2} (\eta_{0}+\eta_{1}) = D \eta_{0},
$$
where $D_{1}$ (resp. $D_{2}$) is the shear rate in the layer of
viscosity $\eta_{0}-\eta_{1}$ (resp. $\eta_{0}+\eta_{1}$). The spatial
average over the configuration is now,
$$
\langle D \rangle = \frac{1}{d} \left[ \frac{d}{2} D_{1} + \frac{d}{2} D_{2}
\right] = \frac{D}{2} \left( \frac{1}{1-\delta} +
\frac{1}{1+\delta} \right) = D \left( 1 + \delta^{2} + \ldots \right),
$$
where $\delta = \eta_{1}/\eta_{0}$. Therefore the average shear rate
across the layer in a parallel configuration always increases with viscosity 
contrast. (In both sets of calculations,
the perpendicular base state is not affected by the shear flow).
Fredrickson's results are based on the introduction of a 
renormalized order parameter mobility that is assumed to 
depend on the average shear rate across the layer. Therefore his 
results for a parallel configuration are affected
by the shear, whereas ours are not. As discussed in Section \ref{sec:lamellae},
the composition field of the base parallel configuration is not modified
by the flow correction arising from viscosity contrast. Finally,
although the implications of the choice of boundary conditions on the
dynamical evolution of partially ordered lamellar configurations are difficult 
to establish, we note that Fredrickson's work focused on fluctuations 
near a uniform, disordered state, whereas we have considered perturbations 
around a (weakly) nonlinear state of a finite amplitude, saturated lamellar 
structure. 

\section*{Acknowledgments} This research has been supported by the National 
Science Council of Taiwan, and also the National Science Foundation
under contract DMR-0100903.

\bibliographystyle{achemso}
\bibliography{$HOME/mss/references}

\begin{thebibliography}{10}
\bibitem{fo:peilong4_1}
The nonlinear saturation of this perturbation will not be addressed, and
  therefore we do need to consider the relative magnitude of the two small
  parameters: $\epsilon$ and the amplitude of the long wavevelength
  perturbation.
\end{thebibliography}

\newpage

\newpage

\newpage

\newpage
\begin{figure}[t]
\putgraph{15cm}{q23reg_g1.eps}
\caption{Stability regions in the $q_1$--$q_2$ and $q_1$--$q_3$ planes 
         with $\omega=0.01$, $\gamma=1$, and $\epsilon=0.04$.}
\label{fig:q23reg}
\end{figure}

\newpage

\end{document}